\documentclass[prd,showpacs,showkeys,floatfix,twocolumn,amsmath,amssymb,floatfix]{revtex4}
\usepackage{graphicx,color,dcolumn,booktabs,bm}
\usepackage{subfigure}
\usepackage{longtable,lscape}
\usepackage{amssymb}
\usepackage{indentfirst}
\usepackage{epsfig}
\usepackage{feynmf}   %{feynmp}
\usepackage{epstopdf}   %{feynmp}
\usepackage{slashed}  %for Feynman symbols
\usepackage{cases}
\usepackage{color}
\usepackage{multirow}
\usepackage{graphicx,color,dcolumn,booktabs,bm}
\usepackage[colorlinks, citecolor=blue,anchorcolor=red,menucolor=red, linkcolor=red,filecolor=red,runcolor=red,urlcolor=blue,frenchlinks=red]{hyperref}

\begin{document}
%=====================================================================================
%=====================================================================================
\title{QCD sum rule studies on the $s s \bar s \bar s$ tetraquark states with $J^{PC} = 1^{+-}$}
%=====================================================================================
%=====================================================================================
%
\author{Er-Liang Cui$^1$}
\author{Hui-Min Yang$^1$}
\author{Hua-Xing Chen$^1$}
\email{hxchen@buaa.edu.cn}
\author{Wei Chen$^2$}
\email{chenwei29@mail.sysu.edu.cn}
\author{Cheng-Ping Shen$^1$}
\email{shencp@ihep.ac.cn}
\affiliation{
$^1$School of Physics, Beihang University, Beijing 100191, China \\
$^2$School of Physics, Sun Yat-Sen University, Guangzhou 510275, China
%$^8$School of Physics and State Key Laboratory of Nuclear Physics and Technology, Peking University, Beijing 100871, China \\
%$^9$Collaborative Innovation Center of Quantum Matter, Beijing 100871, China \\
%$^{10}$Center of High Energy Physics, Peking University, Beijing 100871, China
}
\begin{abstract}
We apply the method of QCD sum rules to study the structure $X$ newly observed by the BESIII Collaboration in the $\phi \eta^\prime$ mass spectrum in 2.0-2.1 GeV region in the $J/\psi \rightarrow \phi \eta \eta^\prime$ decay. We construct all the $s s \bar s \bar s$ tetraquark currents with $J^{PC} = 1^{+-}$, and use them to perform QCD sum rule analyses. One current leads to reliable QCD sum rule results and the mass is extracted to be $2.00^{+0.10}_{-0.09}$ GeV, suggesting that the structure $X$ can be interpreted as an $s s \bar s \bar s$ tetraquark state with $J^{PC} = 1^{+-}$. The $Y(2175)$ can be interpreted as its $s s \bar s \bar s$ partner having $J^{PC} = 1^{--}$, and we propose to search for the other two partners, the $s s \bar s \bar s$ tetraquark states with $J^{PC} = 1^{++}$ and $1^{-+}$, in the $\eta^\prime f_0(980)$, $\eta^\prime K \bar K$, and $\eta^\prime K \bar K^*$ mass spectra.
\end{abstract}
\maketitle
\pagenumbering{arabic}
%
%
%
%=====================================================================================
%=====================================================================================
\section{Introduction}\label{sec:intro}
%=====================================================================================
%=====================================================================================
%

Recently, the BESIII Collaboration reported their observation of a new structure $X$ in the $\phi \eta^\prime$ mass spectrum in 2.0-2.1 GeV region, when studying the $J/\psi \rightarrow \phi \eta \eta^\prime$ decay~\cite{BESIII}. This experiment gives two possibilities:
\begin{enumerate}

\item After assuming $X$ to have the spin-parity quantum numbers $J^P = 1^-$, its mass and decay width are determined to be
\begin{eqnarray}
M_{1^-} &=& 2002.1 \pm 27.5 \pm 15.0~{\rm MeV} \, ,
\label{eq:mass1}
\\ \nonumber \Gamma_{1^-} &=& 129 \pm 17 \pm 7~{\rm MeV} \, .
\end{eqnarray}

\item After assuming $X$ to have the spin-parity quantum numbers $J^P = 1^+$, its mass and decay width are determined to be
\begin{eqnarray}
M_{1^+} &=& 2062.8 \pm 13.1 \pm 4.2~{\rm MeV} \, ,
\label{eq:mass2}
\\ \nonumber \Gamma_{1^+} &=& 177 \pm 36 \pm 20~{\rm MeV} \, .
\end{eqnarray}

\end{enumerate}
Here, the first uncertainties are statistical and the second systematic. The significances are $5.3\sigma$ and $4.9\sigma$, respectively, so these two assumptions can not be distinguished at BESIII. One possible theoretical explanation is to interpret it as an isoscalar axial-vector meson with $I(J^P) = 0(1^+)$, the second radial excitation of $h_1(1380)$~\cite{liu}.

Because the structure $X$ was observed in the $\phi \eta^\prime$ mass spectrum but not reported in the $\phi \eta$ mass spectrum~\cite{BESIII}, it may contain large $\bar s s \bar s s$ component. This makes it a good candidate of exotic hadrons in the light sector~\cite{pdg,Chen:2016qju,Klempt:2007cp,Lebed:2016hpi,Esposito:2016noz,Guo:2017jvc,Olsen:2017bmm}. Another similar candidate is the $Y(2175)$, which was first observed by the BaBar Collaboration in the $\phi f_0(980)$ invariant mass spectrum~\cite{Aubert:2006bu,Aubert:2007ur,Aubert:2007ym,Lees:2011zi}, and later confirmed in the BESII~\cite{Ablikim:2007ab}, BESIII~\cite{Ablikim:2014pfc,Ablikim:2017auj}, and Belle~\cite{Shen:2009zze} experiments. The $Y(2175)$ may also contain large $\bar s s \bar s s$ component, but its measured mass and width are significantly different from those of $X$~\cite{BESIII}.

In our previous studies~\cite{Chen:2008ej,Chen:2018kuu} we have applied the method of QCD sum rules to systematically study the $s s \bar s \bar s$ tetraquark states with $J^{PC} = 1^{--}$. There we found two independent $s s \bar s \bar s$ tetraquark currents with $J^{PC} = 1^{--}$, and the masses are evaluated to be $2.34 \pm 0.17$ GeV and $2.41 \pm 0.25$ GeV, not far from each other~\cite{Chen:2018kuu}. These two values are both significantly larger than the first mass value listed in Eq.~(\ref{eq:mass1}), suggesting that the structure $X$ is difficult to be interpreted as an $s s \bar s \bar s$ tetraquark state of $J^{PC} = 1^{--}$. Instead, the $Y(2175)$ can be well interpreted as an $s s \bar s \bar s$ tetraquark state of $J^{PC} = 1^{--}$~\cite{Chen:2008ej,Chen:2018kuu}. Moreover, the above two mass values are extracted from two diagonalized currents, which do not strongly correlate to each other and may couple to two different physical states: one is the $Y(2175)$, and the other is around 2.4 GeV. There have been some evidences for the latter structure in the previous experiments~\cite{Aubert:2007ur,Ablikim:2007ab,Shen:2009zze,Ablikim:2014pfc}, and we refer to Ref.~\cite{Chen:2018kuu} for detailed discussions.

In the present study we follow the same approach to study the $s s \bar s \bar s$ tetraquark states with $J^{PC} = 1^{+-}$, and examine whether the structure $X$ can be explained. Again, we shall find that there are two independent $s s \bar s \bar s$ tetraquark currents with $J^{PC} = 1^{+-}$, which we shall use to perform QCD sum rule analyses. The internal structures of exotic hadrons are always complicated. For each internal structure we can construct the relevant interpolating current, and there are usually many interpolating currents when studying multiquark states. In this case, the only two independent currents make it possible to study their mixing. Note that we have done this in Ref.~\cite{Chen:2018kuu} when studying the $s s \bar s \bar s$ tetraquark states with $J^{PC} = 1^{--}$. By doing this we can carefully examine the relations between physical states and the relevant interpolating currents, and further understand the internal structures of exotic hadrons.

This paper is organized as follows. In Sec.~\ref{sec:current}, we systematically construct the $s s \bar s \bar s$ tetraquark currents with $J^{PC} = 1^{+-}$, using both diquark/antidiquark fields and quark-antiquark pairs. These currents are then used to perform QCD sum rule analyses in Sec.~\ref{sec:sumrule}, and numerical analyses in Sec.~\ref{sec:numerical}. Their mixing are investigated in Sec.~\ref{sec:mixing}. Sec.~\ref{sec:summary} is a summary.

%
%=====================================================================================
%=====================================================================================
\section{Interpolating Currents}\label{sec:current}
%=====================================================================================
%=====================================================================================
%

The $s s \bar s \bar s$ tetraquark currents with the quantum numbers $J^{PC} = 1^{--}$ have been systematically constructed in Ref.~\cite{Chen:2008ej}. See also Refs.~\cite{Chen:2008qw,Chen:2008ne,Chen:2013jra} where many other vector and axial-vector tetraquark currents are systematically constructed. In this section we follow the same approach to construct the $s s \bar s \bar s$ tetraquark currents with the quantum numbers $J^{PC} = 1^{+-}$. We find two non-vanishing diquark-antidiquark currents:
%
%%%%%%%%%%%%%%%%%%%%%%%%%%%%%%%%%%%%%%%%%%%%%%%%%%%%%%%%%%%%%%%%%%%%%%%%%%%%%%
\begin{eqnarray}
%-------------------------------------eta 1------------------------------------
\eta_{1\mu} &=& (s_a^T C s_b) (\bar{s}_a \gamma_\mu \gamma_5 C \bar{s}_b^T)
\label{def:eta1}
\\ \nonumber && ~~~~~~~~~~~~~  - (s_a^T C \gamma_\mu \gamma_5 s_b) (\bar{s}_a C \bar{s}_b^T)  \, ,
%-------------------------------------eta 2------------------------------------
\\ \eta_{2\mu} &=& (s_a^T C \gamma^\nu s_b) (\bar{s}_a \sigma_{\mu\nu} \gamma_5 C \bar{s}_b^T)
\label{def:eta2}
\\ \nonumber && ~~~~~~~~~~~~~  - (s_a^T C \sigma_{\mu\nu} \gamma_5 s_b) (\bar{s}_a \gamma^\nu C \bar{s}_b^T) \, ,
\end{eqnarray}
%%%%%%%%%%%%%%%%%%%%%%%%%%%%%%%%%%%%%%%%%%%%%%%%%%%%%%%%%%%%%%%%%%%%%%%%%%%%%%
%
where $a$ and $b$ are color indices; $C = i\gamma_2 \gamma_0$ is the charge-conjugation operator; the sum over repeated indices is taken.
These two diquark-antidiquark currents are independent of each other. Recalling that the diquark fields $s_a^T C s_b/s_a^T C \gamma_\mu \gamma_5 s_b/s_a^T C \gamma_\mu s_b/s_a^T C \sigma_{\mu\nu} \gamma_5 s_b$ have the quantum numbers $J^P = 0^-/1^-/1^+/1^\pm$, respectively, the first current $\eta_{1\mu}$ only contains excited diquark and antidiquark fields, but the second one $\eta_{2\mu}$ contains (at least) one ground-state diquark/antidiquark field. Hence, $\eta_{2\mu}$ has a more stable internal structure and may lead to better sum rule results.

Besides the above diquark-antidiquark currents, we find that there are four mesonic-mesonic currents:
%
%%%%%%%%%%%%%%%%%%%%%%%%%%%%%%%%%%%%%%%%%%%%%%%%%%%%%%%%%%%%%%%%%%%%%%%%%%%%%%
\begin{eqnarray} \nonumber
%-------------------------------------eta 3------------------------------------
\eta_{3\mu} &=& (\bar{s}_a \gamma_5 s_a)(\bar{s}_b \gamma_\mu s_b) \, ,
%-------------------------------------eta 4------------------------------------
\\ \nonumber \eta_{4\mu} &=& (\bar{s}_a \gamma^\nu \gamma_5 s_a)(\bar{s}_b \sigma_{\mu\nu} s_b) \, ,
\\ \nonumber \eta_{5\mu} &=& {\lambda_{ab}}{\lambda_{cd}}(\bar{s}_a \gamma_5 s_b)(\bar{s}_c \gamma_\mu s_d) \, ,
%-------------------------------------eta 6------------------------------------
\\ \nonumber \eta_{6\mu} &=& {\lambda_{ab}}{\lambda_{cd}} (\bar{s}_a \gamma^\nu\gamma_5 s_b)(\bar{s}_c \sigma_{\mu\nu} s_d) \, .
\end{eqnarray}
%%%%%%%%%%%%%%%%%%%%%%%%%%%%%%%%%%%%%%%%%%%%%%%%%%%%%%%%%%%%%%%%%%%%%%%%%%%%%%
%
The following relations can be verified by using the Fierz transformation, so the number of independent mesonic-mesonic currents is also two:
%
%%%%%%%%%%%%%%%%%%%%%%%%%%%%%%%%%%%%%%%%%%%%%%%%%%%%%%%%%%%%%%%%%%%%%%%%%%%%%%
\begin{eqnarray}
\eta_{5\mu} &=& - \frac{5}{3} \eta_{3\mu} - i \eta_{4\mu} \, ,
\\
\eta_{6\mu} &=& 3i \eta_{3\mu} + \frac{1}{3} \eta_{4\mu} \, .
\end{eqnarray}
%%%%%%%%%%%%%%%%%%%%%%%%%%%%%%%%%%%%%%%%%%%%%%%%%%%%%%%%%%%%%%%%%%%%%%%%%%%%%%
%
Moreover, we can use the Fierz transformation to relate the diquark-antidiquark and mesonic-mesonic currents:
%
%%%%%%%%%%%%%%%%%%%%%%%%%%%%%%%%%%%%%%%%%%%%%%%%%%%%%%%%%%%%%%%%%%%%%%%%%%%%%%
\begin{eqnarray}
\eta_{1\mu} &=& - \eta_{3\mu} + i \eta_{4\mu} \, ,
\\ \eta_{2\mu} &=& 3i \eta_{3\mu} - \eta_{4\mu} \, .
\end{eqnarray}
%%%%%%%%%%%%%%%%%%%%%%%%%%%%%%%%%%%%%%%%%%%%%%%%%%%%%%%%%%%%%%%%%%%%%%%%%%%%%%
%
Therefore, these two constructions are equivalent.

In the following we shall use $\eta_{1\mu}$ and $\eta_{2\mu}$ to perform QCD sum rule analyses.

%
%=====================================================================================
%=====================================================================================
\section{QCD sum rule Analyses}\label{sec:sumrule}
%=====================================================================================
%=====================================================================================
%

QCD sum rules~\cite{Shifman:1978bx,Reinders:1984sr,Nielsen:2009uh}, a powerful and successful non-perturbative method, have been widely applied to study various exotic hadrons~\cite{Chen:2007zzg,Chen:2007xr,Lee:2006vk,Wang:2006ri,Zhang:2006xp,Matheus:2006xi,Matheus:2007ta,Sugiyama:2007sg,Zhang:2011jja,Agaev:2016mjb,Wang:2015epa,Huang:2016rro}.
In this method we calculate the two-point correlation function at both the hadron and quark-gluon levels:
%
%%%%%%%%%%%%%%%%%%%%%%%%%%%%%%%%%%%%%%%%%%%%%%%%%%%%%%%%%%%%%%%%%%%%%%%%%%%%%%
\begin{eqnarray}
\Pi_{\mu\nu}(q^2) &\equiv& i \int d^4x e^{iqx} \langle 0 | T \eta_\mu(x) { \eta_\nu^\dagger } (0) | 0 \rangle
\label{def:pi}
\\ \nonumber &=& ( {q_\mu q_\nu \over q^2} - g_{\mu\nu} ) \Pi(q^2) + {q_\mu q_\nu \over q^2} \Pi^{(0)}(q^2) \, .
\label{def:pi1}
\end{eqnarray}
%%%%%%%%%%%%%%%%%%%%%%%%%%%%%%%%%%%%%%%%%%%%%%%%%%%%%%%%%%%%%%%%%%%%%%%%%%%%%%
%

At the hadron level, we can express $\Pi(q^2)$ in the form of the dispersion relation with a spectral function $\rho(s)$:
%
%%%%%%%%%%%%%%%%%%%%%%%%%%%%%%%%%%%%%%%%%%%%%%%%%%%%%%%%%%%%%%%%%%%%%%%%%%%%%%
\begin{equation}
\Pi(q^2)=\int^\infty_{16 m_s^2}\frac{\rho(s)}{s-q^2-i\varepsilon}ds \, .
\label{eq:disper}
\end{equation}
%%%%%%%%%%%%%%%%%%%%%%%%%%%%%%%%%%%%%%%%%%%%%%%%%%%%%%%%%%%%%%%%%%%%%%%%%%%%%%
%
Then we adopt a parametrization of one pole dominance and a continuum contribution:
%
%%%%%%%%%%%%%%%%%%%%%%%%%%%%%%%%%%%%%%%%%%%%%%%%%%%%%%%%%%%%%%%%%%%%%%%%%%%%%%
\begin{eqnarray}
\rho(s) & \equiv & \sum_n\delta(s-M^2_n)\langle 0|\eta|n\rangle\langle n|{\eta^\dagger}|0\rangle
\label{eq:rho}
\\ \nonumber &=& f^2_X\delta(s-M^2_X)+ \rm{higher\,\,states} \, ,
\end{eqnarray}
%%%%%%%%%%%%%%%%%%%%%%%%%%%%%%%%%%%%%%%%%%%%%%%%%%%%%%%%%%%%%%%%%%%%%%%%%%%%%%
%
where $X$ is the ground state.

At the quark-gluon level, we insert $\eta_{1\mu}$ and $\eta_{2\mu}$ into Eq.~(\ref{def:pi}), which are then calculated using the method of operator product expansion (OPE).
After performing the Borel transformation at both the hadron and quark-gluon levels, we obtain
%
%%%%%%%%%%%%%%%%%%%%%%%%%%%%%%%%%%%%%%%%%%%%%%%%%%%%%%%%%%%%%%%%%%%%%%%%%%%%%%
\begin{equation}
\Pi(M_B^2,\infty)\equiv\mathcal{B}_{M_B^2}\Pi(q^2)=\int^\infty_{16 m_s^2}~e^{-s/M_B^2}~\rho(s)ds \, .
\label{eq_borel}
\end{equation}
%%%%%%%%%%%%%%%%%%%%%%%%%%%%%%%%%%%%%%%%%%%%%%%%%%%%%%%%%%%%%%%%%%%%%%%%%%%%%%
%
Then we approximate the continuum using the spectral density of OPE above a threshold value $s_0$, and obtain the following sum rule equation
%
%%%%%%%%%%%%%%%%%%%%%%%%%%%%%%%%%%%%%%%%%%%%%%%%%%%%%%%%%%%%%%%%%%%%%%%%%%%%%%
\begin{equation}
\Pi(M_B^2, s_0) \equiv f^2_X~e^{-M_X^2/M_B^2} = \int^{s_0}_{16 m_s^2}~e^{-s/M_B^2}~\rho(s)ds
\label{eq_fin} \, .
\end{equation}
%%%%%%%%%%%%%%%%%%%%%%%%%%%%%%%%%%%%%%%%%%%%%%%%%%%%%%%%%%%%%%%%%%%%%%%%%%%%%%
%
Finally, we can use this equation to calculate $M_X$, the mass of the ground state $X$, through
%
%%%%%%%%%%%%%%%%%%%%%%%%%%%%%%%%%%%%%%%%%%%%%%%%%%%%%%%%%%%%%%%%%%%%%%%%%%%%%%
\begin{eqnarray}
M^2_X(M_B^2, s_0) &=& \frac{\frac{\partial}{\partial(-1/M_B^2)}\Pi(M_B^2, s_0)}{\Pi(M_B^2, s_0)}
\label{eq_LSR}
\\ \nonumber &=& \frac{\int^{s_0}_{16 m_s^2}~e^{-s/M_B^2}~s~\rho(s)ds}{\int^{s_0}_{16 m_s^2}~e^{-s/M_B^2}~\rho(s)ds} \, .
\end{eqnarray}
%%%%%%%%%%%%%%%%%%%%%%%%%%%%%%%%%%%%%%%%%%%%%%%%%%%%%%%%%%%%%%%%%%%%%%%%%%%%%%
%

%
%%%%%%%%%%%%%%%%%%%%%%%%%%%%%%%%%%%%%%%%%%%%%%%%%%%%%%%%%%%%%%%%%%%%%%%%%%%%%%
\begin{figure*}[hbt]
\normalsize
\hrulefill
\begin{eqnarray}
%%%%%%%%%%%%%%%%%%%%%%%%%%%%%%%%%%%%%%%%%%%%%%%%%%%%%%%%%%%%%%%%%%%%%%%%%%%%%%
%------------------------------\rho 1-- eta_1----------------------------------
\label{eq:pieta1}
\Pi_{\eta_1\eta_1} &=& \int^{s_0}_{16 m_s^2} \Bigg [
{s^4 \over 18432 \pi^6}
- { 5 m_s^2 s^3 \over 768 \pi^6 }
+ \Big ( - { \langle g^2 G G \rangle \over 18432 \pi^6 }
+ {5 m_s \langle \bar s s \rangle \over 48 \pi^4} \Big ) s^2
\\ \nonumber &&
+ \Big ( - { 5 \langle \bar s s \rangle^2 \over 18 \pi^2 }
+ { 35 m_s \langle g \bar s \sigma G s \rangle \over 576 \pi^4 }
+ { 17 m_s^2 \langle g^2 G G \rangle \over 4608 \pi^6 } \Big ) s
\\ \nonumber &&
+ \Big ( - { 7 \langle \bar s s \rangle \langle g \bar s \sigma G s \rangle \over 48 \pi^2 }
- { m_s \langle g^2 G G \rangle \langle \bar s s \rangle \over 64 \pi^4}
+ { 17 m_s^2 \langle \bar s s \rangle^2 \over 4 \pi^2 } \Big ) \Bigg ] e^{-s/M_B^2} ds
\\ \nonumber &&
+ \Big (  { \langle g^2 GG \rangle \langle \bar s s \rangle^2 \over 144 \pi^2}
- { \langle g \bar s \sigma G s \rangle^2 \over 288 \pi^2 }
- {20 m_s \langle \bar s s \rangle^3 \over 9}
- { m_s \langle g^2 GG \rangle \langle g \bar s \sigma G s \rangle \over 384 \pi^4 }
+ { 67 m_s^2 \langle \bar s s \rangle \langle g \bar s \sigma G s \rangle \over 48 \pi^2 } \Big )
\\ \nonumber &&
+ {1 \over M_B^2} \Big (  { 32 g^2 \langle \bar s s \rangle^4 \over 81 }
- { \langle g^2 GG \rangle \langle \bar s s \rangle \langle g \bar s \sigma G s \rangle \over 96 \pi^2 }
+ { 67 m_s \langle \bar s s \rangle^2 \langle g \bar s \sigma G s \rangle \over 36 }
+ { m_s^2 \langle g^2 GG \rangle \langle \bar s s \rangle^2 \over 576 \pi^2 }
- { 19 m_s^2 \langle g \bar s \sigma G s \rangle^2 \over 96 \pi^2 } \Big )
\, ,
%------------------------------\rho 2-- eta_2----------------------------------
\\ \label{eq:pieta2}
\Pi_{\eta_2\eta_2} &=& \int^{s_0}_{16 m_s^2} \Bigg [
{s^4 \over 12288 \pi^6}
- { m_s^2 s^3 \over 2560 \pi^6 }
+ \Big ( { \langle g^2 G G \rangle \over 18432 \pi^6 }
- {13 m_s \langle \bar s s \rangle \over 96 \pi^4} \Big ) s^2
\\ \nonumber &&
+ \Big ( { 25 \langle \bar s s \rangle^2 \over 36 \pi^2 }
- { 155 m_s \langle g \bar s \sigma G s \rangle \over 576 \pi^4 }
- { m_s^2 \langle g^2 G G \rangle \over 2304 \pi^6 } \Big ) s
+ \Big ( { 31 \langle \bar s s \rangle \langle g \bar s \sigma G s \rangle \over 48 \pi^2 }
- { 13 m_s^2 \langle \bar s s \rangle^2 \over 8 \pi^2 }  \Big ) \Bigg ] e^{-s/M_B^2} ds
\\ \nonumber &&
+ \Big (
 { \langle g \bar s \sigma G s \rangle^2 \over 18 \pi^2 }
- {14 m_s \langle \bar s s \rangle^3 \over 9}
+ { 11 m_s^2 \langle \bar s s \rangle \langle g \bar s \sigma G s \rangle \over 48 \pi^2 } \Big )
\\ \nonumber &&
+ {1 \over M_B^2} \Big ( { 16 g^2 \langle \bar s s \rangle^4 \over 27 }
+ { 11 m_s \langle \bar s s \rangle^2 \langle g \bar s \sigma G s \rangle \over 36 }
- { m_s^2 \langle g^2 GG \rangle \langle \bar s s \rangle^2 \over 576 \pi^2 }
+ { 13 m_s^2 \langle g \bar s \sigma G s \rangle^2 \over 96 \pi^2 } \Big )
\, ,
\\ \label{eq:eta12}
\Pi_{\eta_1\eta_2} &=& i \int^{s_0}_{16 m_s^2} \Bigg [
{ \langle g^2 G G \rangle \over 6144 \pi^6 } s^2
+ \Big ( - { 5 m_s \langle g \bar s \sigma G s \rangle \over 192 \pi^4 }
- { m_s^2 \langle g^2 G G \rangle \over 768 \pi^6 } \Big ) s
+ { \langle \bar s s \rangle \langle g \bar s \sigma G s \rangle \over 16 \pi^2 } \Bigg ] e^{-s/M_B^2} ds
\\ \nonumber &&
+ \Big (
 { \langle g \bar s \sigma G s \rangle^2 \over 96 \pi^2 }
- { m_s^2 \langle \bar s s \rangle \langle g \bar s \sigma G s \rangle \over 16 \pi^2 } \Big )
+ {1 \over M_B^2} \Big (
- { m_s \langle \bar s s \rangle^2 \langle g \bar s \sigma G s \rangle \over 12 }
- { m_s^2 \langle g^2 GG \rangle \langle \bar s s \rangle^2 \over 192 \pi^2 }
+ { m_s^2 \langle g \bar s \sigma G s \rangle^2 \over 32 \pi^2 } \Big )
\, .
\end{eqnarray}
\hrulefill
\vspace*{4pt}
\end{figure*}
%%%%%%%%%%%%%%%%%%%%%%%%%%%%%%%%%%%%%%%%%%%%%%%%%%%%%%%%%%%%%%%%%%%%%%%%%%%%%%

For the currents $\eta_{1\mu}$ and $\eta_{2\mu}$, we have calculated the OPE up to dimension twelve. Explicitly, we have calculated the perturbative term, the gluon condensate $\langle g_s^2 GG \rangle$, the quark condensate $\langle \bar s s \rangle$, the quark-gluon condensate $\langle g_s \bar s \sigma G s \rangle$, and their combinations $\langle g_s^2 GG \rangle \langle \bar s s \rangle$, $\langle g_s^2 GG \rangle \langle \bar s s \rangle^2$, $\langle g_s^2 GG \rangle \langle g_s \bar s \sigma G s \rangle$, $\langle g_s^2 GG \rangle \langle \bar s s \rangle \langle g_s \bar s \sigma G s \rangle$, $\langle \bar s s \rangle^2$, $\langle \bar s s \rangle^3$, $\langle \bar s s \rangle^4$, $\langle g_s \bar s \sigma G s \rangle^2$, $\langle \bar s s \rangle \langle g_s \bar s \sigma G s \rangle$, and $\langle \bar s s \rangle^2 \langle g_s \bar s \sigma G s \rangle$. The results for $\eta_{1\mu}$ and $\eta_{2\mu}$ are shown in Eqs.~(\ref{eq:pieta1}) and (\ref{eq:pieta2}), respectively. For completeness, we have also calculated the sum rules for the off-diagonal term:
%
%%%%%%%%%%%%%%%%%%%%%%%%%%%%%%%%%%%%%%%%%%%%%%%%%%%%%%%%%%%%%%%%%%%%%%%%%%%%%%
\begin{eqnarray}
\Pi_{\mu\nu}^{\eta_1\eta_2}(q^2) &=& i \int d^4x e^{iqx} \langle 0 | T \eta_{1\mu}(x) { \eta_{2\nu}^\dagger } (0) | 0 \rangle
\label{eq:offdiagonal}
\\ \nonumber &=& ( {q_\mu q_\nu \over q^2} - g_{\mu\nu} ) \Pi_{\eta_1\eta_2}(q^2) + {q_\mu q_\nu \over q^2} \Pi^{(0)}_{\eta_1\eta_2}(q^2) \, .
\end{eqnarray}
%%%%%%%%%%%%%%%%%%%%%%%%%%%%%%%%%%%%%%%%%%%%%%%%%%%%%%%%%%%%%%%%%%%%%%%%%%%%%%
%
After performing the Borel transformation to $\Pi_{\eta_1\eta_2}(q^2)$, we obtain $\Pi_{\eta_1\eta_2}(M_B^2)$ whose explicit expression is shown in Eq.~(\ref{eq:eta12}).

%
%=====================================================================================
%=====================================================================================
\section{Numerical Analyses}\label{sec:numerical}
%=====================================================================================
%=====================================================================================
%

In this section we use the currents $\eta_{1\mu}$ and $\eta_{2\mu}$ to perform numerical analyses, for which we use the following values for various condensates~\cite{pdg,Yang:1993bp,Narison:2002pw,Gimenez:2005nt,Jamin:2002ev,Ioffe:2002be,Ovchinnikov:1988gk,Ellis:1996xc}:
%
%%%%%%%%%%%%%%%%%%%%%%%%%%%%%%%%%%%%%%%%%%%%%%%%%%%%%%%%%%%%%%%%%%%%%%%%%%%%%%
\begin{eqnarray}
\nonumber &&m_s(2~\mbox{GeV})= 96 ^{+8}_{-4} \mbox{ MeV}\, ,
\\ \nonumber && \alpha_s(1.7~\mbox{GeV}) = 0.328 \pm 0.03 \pm 0.025 \, ,
\\ \nonumber &&\langle\bar qq \rangle=-(0.24 \pm 0.01 \mbox{ GeV})^3\, ,
\\ \label{condensates} &&\langle\bar ss\rangle=-(0.8\pm 0.1)\times(0.240 \mbox{ GeV})^3\, ,
\\ \nonumber &&\langle g_s^2GG\rangle =(0.48\pm 0.14) \mbox{ GeV}^4\, ,
\\ \nonumber && \langle g_s\bar q\sigma G q\rangle=-M_0^2\times\langle\bar qq\rangle\, ,
\\ \nonumber && M_0^2=(0.8\pm0.2)\mbox{ GeV}^2\, .
\end{eqnarray}
%%%%%%%%%%%%%%%%%%%%%%%%%%%%%%%%%%%%%%%%%%%%%%%%%%%%%%%%%%%%%%%%%%%%%%%%%%%%%%
%

%
%%%%%%%%%%%%%%%%%%%%%%%%%%%%%%%%%%%%%%%%%%%%%%%%%%%%%%%%%%%%%%%%%%%%%%%%%%%%%%
%---------figure current 1
\begin{figure*}[hbt]
\begin{center}
\includegraphics[width=0.4\textwidth]{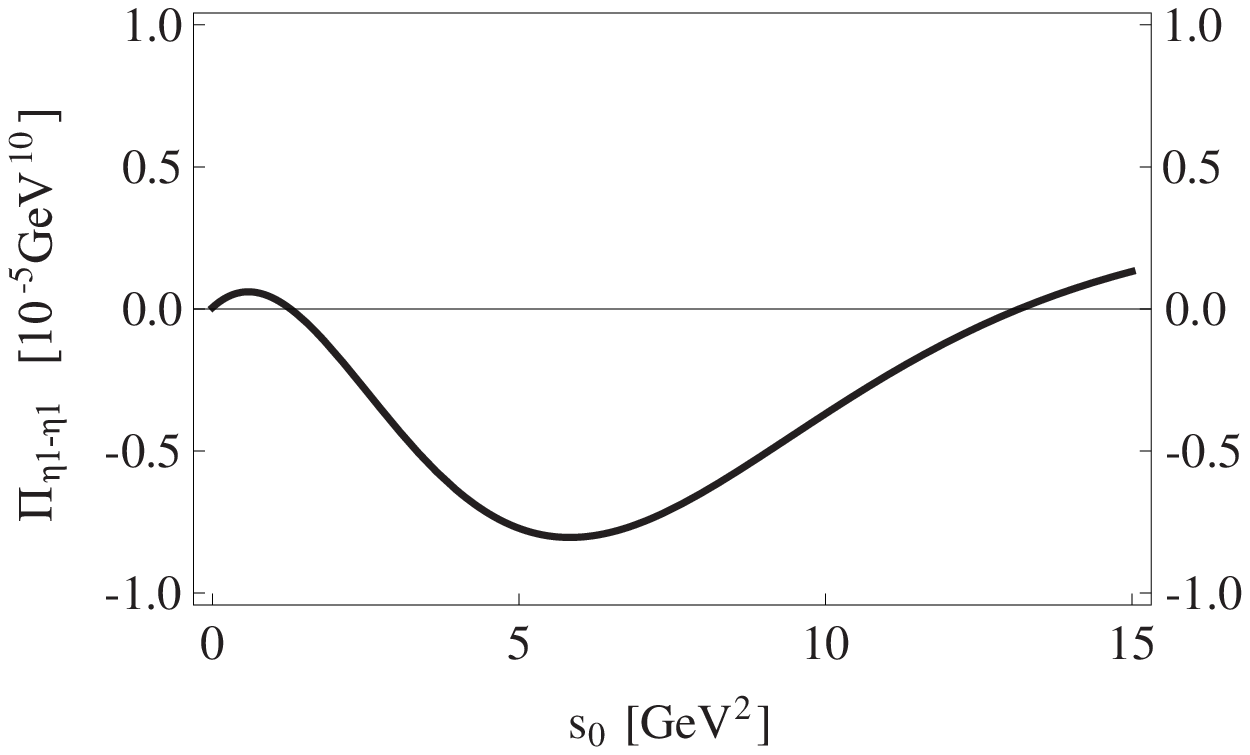}
\includegraphics[width=0.4\textwidth]{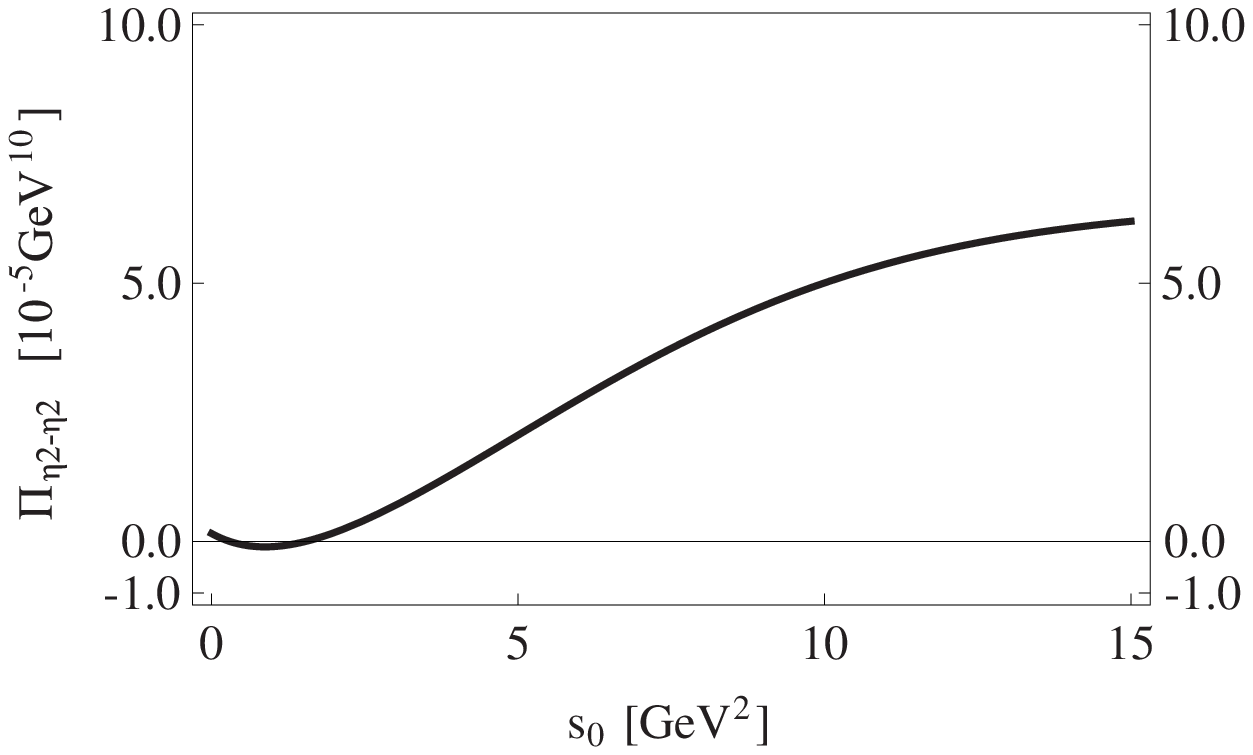}
\caption{The correlation functions $\Pi_{\eta_1\eta_1}(M_B^2, s_0)$ and $\Pi_{\eta_2\eta_2}(M_B^2, s_0)$ as functions of the threshold value $s_0$. The curves are obtained by taking $M_B^2 = 1.8$ GeV$^2$.}
\label{fig:pi}
\end{center}
\end{figure*}
%%%%%%%%%%%%%%%%%%%%%%%%%%%%%%%%%%%%%%%%%%%%%%%%%%%%%%%%%%%%%%%%%%%%%%%%%%%%%%
%

Different from Ref.~\cite{Chen:2008ej} where there are two independent $s s \bar s \bar s$ tetraquark currents with $J^{PC} = 1^{--}$ leading to similar QCD sum rule results, in the present study we find that the two $s s \bar s \bar s$ tetraquark currents with $J^{PC} = 1^{+-}$, $\eta_{1\mu}$ and $\eta_{2\mu}$, lead to totally different sum rule results. This can be clearly seen in Fig.~\ref{fig:pi}, where we show the Borel transformed correlation functions $\Pi_{\eta_1\eta_1}(M_B^2, s_0)$ and $\Pi_{\eta_2\eta_2}(M_B^2, s_0)$ as functions of the threshold value $s_0$. We find that $\Pi_{\eta_1\eta_1}(M_B^2, s_0)$ is negative, and so non-physical, in the region $s_0< 10$~GeV$^2$. Hence, it can not strongly couple to any structure that is smaller than 3.0~GeV. The situation for $\eta_{2\mu}$ is different since $\Pi_{\eta_2\eta_2}(M_B^2, s_0)$ is positive and well defined. This behavior seems to be reasonable because $\eta_{1\mu}$ only contains excited diquark and antidiquark fields, while $\eta_{2\mu}$ contains (at least) one ground-state diquark/antidiquark field and so more stable.

\begin{figure*}[hbt]
\begin{center}

\includegraphics[width=0.4\textwidth]{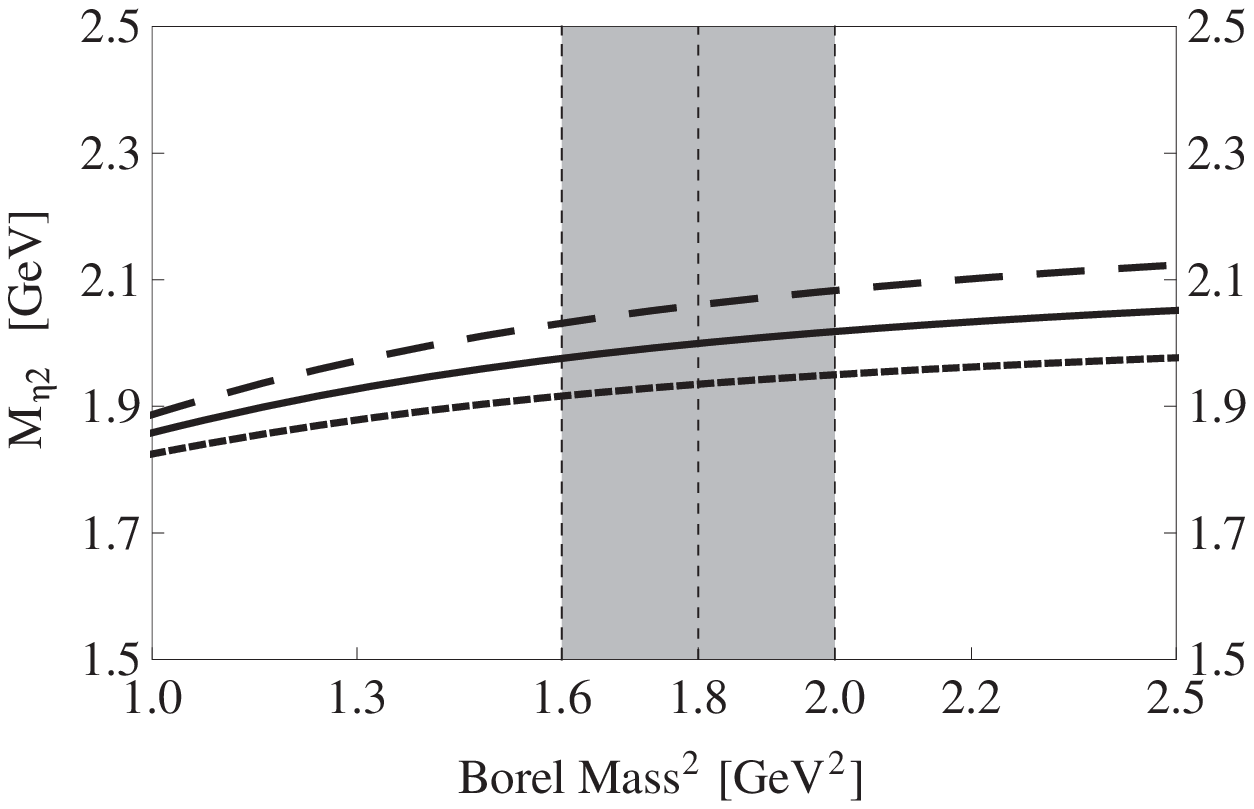}
\includegraphics[width=0.4\textwidth]{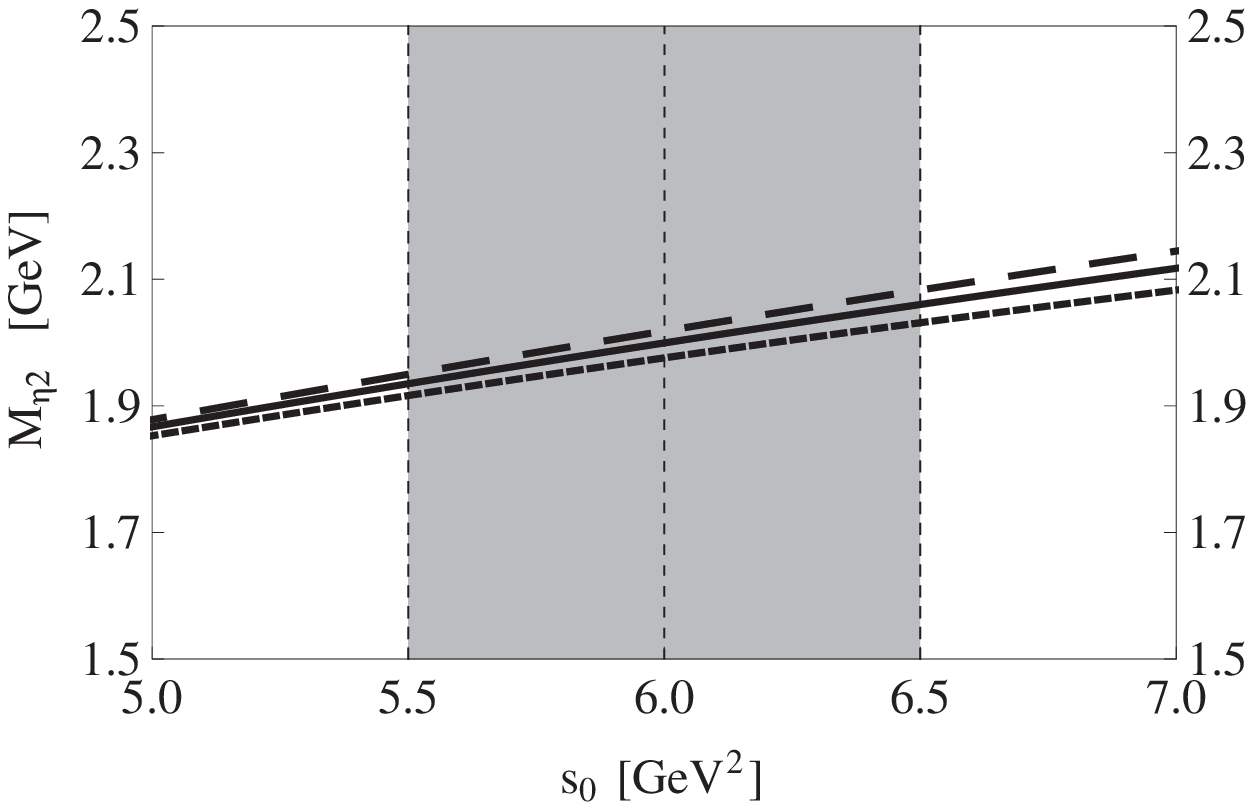}
\caption{
The mass extracted from the current $\eta_{2\mu}$, denoted as $M_{\eta_2}$, as a function of the Borel mass $M_B$ (left) and the threshold value $s_0$ (right).
In the left panel, the short-dashed/solid/long-dashed curves are obtained by setting $s_0 = 5.5/6.0/6.5$ GeV$^2$, respectively.
In the right panel, the short-dashed/solid/long-dashed curves are obtained by setting $M_B^2 = 1.6/1.8/2.0$ GeV$^2$, respectively.}
\label{fig:eta2}
\end{center}
\end{figure*}

In the following we shall only use the current $\eta_{2\mu}$ to perform numerical analyses. After carefully investigating a) the OPE convergence, b) the pole contribution, and c) the mass dependence on the two free parameters $M_B$ and $s_0$, we obtain reliable QCD sum rule results in the regions $1.6$~GeV$^2 < M_B^2< 2.0$~GeV$^2$ and $5.5$~GeV$^2 < s_0< 6.5$~GeV$^2$:
\begin{itemize}

\item First we study the convergence of the operator product expansion. After taking $s_0$ to be $\infty$ and the integral subscript $16 m_s^2$ to be zero, we obtain the numerical series of
the OPE as a function of $M_B$:
\begin{eqnarray}
&& \Pi_{\eta_2\eta_2}(M_B^2, \infty) =
\\ \nonumber && ~~ + 2.0 \times 10^{-6} M_B^{10} - 2.2 \times 10^{-8} M_B^8 + 3.0 \times 10^{-6} M_B^6
\\ \nonumber && ~~ + 6.1 \times 10^{-6} M_B^4 - 6.5 \times 10^{-6} M_B^2
\\ \nonumber && ~~ + 6.2 \times 10^{-7}  M_B^{0} + 7.5 \times 10^{-8} M_B^{-2} \, .
\end{eqnarray}
From this equation, we clearly see that the OPE convergence is quite good: the dimension 12 terms ($\sim M_B^{-2}$) are significantly smaller than the dimension 10 terms ($\sim M_B^{0}$), which are again significantly smaller than the dimension 8 terms ($\sim M_B^{2}$). Numerically, we show the ratio
\begin{eqnarray}
\mbox{CVG} \equiv {\Pi^{{\rm Dim}=10+12}_{\eta_2\eta_2}(M_B^2, s_0) \over \Pi_{\eta_2\eta_2}(M_B^2, s_0)} \, ,
\label{eq:cvg}
\end{eqnarray}
in Fig.~\ref{fig:cvg} as a function of the Borel mass $M_B$. We find it to be smaller than 5\% in the regions $1.6$~GeV$^2 < M_B^2< 2.0$~GeV$^2$ and $5.5$~GeV$^2 < s_0< 6.5$~GeV$^2$.

%
%%%%%%%%%%%%%%%%%%%%%%%%%%%%%%%%%%%%%%%%%%%%%%%%%%%%%%%%%%%%%%%%%%%%%%%%%%%%%%
%---------figure current 1
\begin{figure}[!hbt]
\begin{center}
\includegraphics[width=0.4\textwidth]{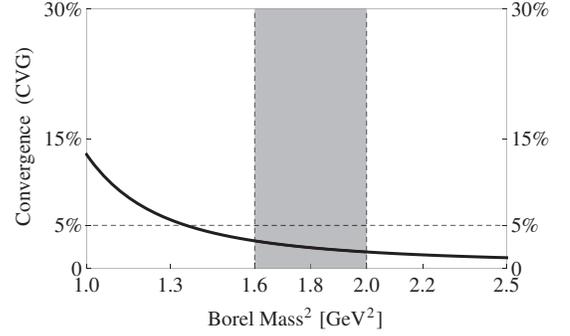}
\caption{The ratio CVG, defined in Eq.~(\ref{eq:cvg}), as a function of the Borel mass $M_B$. The curve is obtained by taking $s_0 = 6.0$ GeV$^2$.}
\label{fig:cvg}
\end{center}
\end{figure}
%%%%%%%%%%%%%%%%%%%%%%%%%%%%%%%%%%%%%%%%%%%%%%%%%%%%%%%%%%%%%%%%%%%%%%%%%%%%%%
%

\item Then we study the pole contribution, defined as
\begin{eqnarray}
\mbox{PC} \equiv {\Pi_{\eta_2\eta_2}(M_B^2, s_0) \over \Pi_{\eta_2\eta_2}(M_B^2, \infty)} \, .
\label{eq:pc}
\end{eqnarray}
We show it as a function of the Borel mass $M_B$ in Fig.~\ref{fig:pc}.
We find it to be 30\% $<$ PC $<$ 58\% in the regions $1.6$~GeV$^2 < M_B^2< 2.0$~GeV$^2$ and $5.5$~GeV$^2 < s_0< 6.5$~GeV$^2$.
This amount of pole contribution is acceptable when one applies the method of QCD sum rules to study multiquark states.

%
%%%%%%%%%%%%%%%%%%%%%%%%%%%%%%%%%%%%%%%%%%%%%%%%%%%%%%%%%%%%%%%%%%%%%%%%%%%%%%
%---------figure current 1
\begin{figure}[!hbt]
\begin{center}
\includegraphics[width=0.4\textwidth]{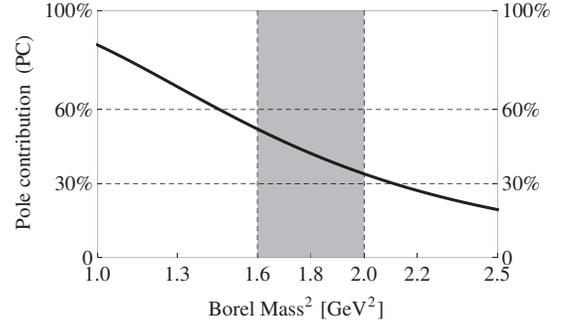}
\caption{The pole contribution (PC), defined in Eq.~(\ref{eq:pc}), as a function of the Borel mass $M_B$. The curve is obtained by taking $s_0 = 6.0$ GeV$^2$.}
\label{fig:pc}
\end{center}
\end{figure}
%%%%%%%%%%%%%%%%%%%%%%%%%%%%%%%%%%%%%%%%%%%%%%%%%%%%%%%%%%%%%%%%%%%%%%%%%%%%%%
%

\item Finally we study the mass dependence on the two free parameters, the Borel mass $M_B$ and the threshold value $s_0$. To clearly see this, we show $M_{\eta_2}$, the mass extracted from the current $\eta_{2\mu}$, in Fig.~\ref{fig:eta2} as a function of $M_B$ and $s_0$.

    In the left panel we show $M_{\eta_2}$ as a function of the Borel mass $M_B$, and find it quite stable in the Borel window $1.6$~GeV$^2 < M_B^2< 2.0$~GeV$^2$. Comparing this figure with Fig.~\ref{fig:cvg} and Fig.~\ref{fig:pc}, we find that one can obtain a still larger pole contribution by choosing a smaller Borel mass (as shown in Fig.~\ref{fig:pc}), but at the same time the convergence of OPE would become worse (as shown in Fig.~\ref{fig:cvg}) and the mass dependence on the Borel mass would become stronger (as shown in the left panel of Fig.~\ref{fig:eta2}). Considering all these behaviours, we find it suitable to fix the Borel window to be $1.6$~GeV$^2 < M_B^2< 2.0$~GeV$^2$.

    In the right panel we show $M_{\eta_2}$ as a function of the threshold value $s_0$. We find that the mass curves moderately depend on the threshold value $s_0$. Especially, we evaluate the mass to be $1.94$~GeV$<M_{\eta_2}<2.06$~GeV in the region $5.5$~GeV$^2 < s_0< 6.5$~GeV$^2$. This uncertainty is about 6\%, quite typical in QCD sum rule studies.

\end{itemize}

%
%%%%%%%%%%%%%%%%%%%%%%%%%%%%%%%%%%%%%%%%%%%%%%%%%%%%%%%%%%%%%%%%%%%%%%%%%%%%%%
%---------figure current 1
%%%%%%%%%%%%%%%%%%%%%%%%%%%%%%%%%%%%%%%%%%%%%%%%%%%%%%%%%%%%%%%%%%%%%%%%%%%%%%
%

Summarizing the above analyses, we have used the $s s \bar s \bar s$ tetraquark current $\eta_{2\mu}$ with $J^{PC} = 1^{+-}$ to perform QCD sum rule analyses. After carefully choosing the working regions to be $1.6$~GeV$^2 < M_B^2< 2.0$~GeV$^2$ and $5.5$~GeV$^2 < s_0< 6.5$~GeV$^2$, we extract the mass to be
\begin{eqnarray}
M_{\eta_2} &=& 2.00~^{+0.02}_{-0.02}~^{+0.06}_{-0.06}~^{+0.07}_{-0.06} {\rm~GeV}
\label{eq:mass}
\\ \nonumber &=& 2.00 ^{+0.10}_{-0.09} {\rm~GeV} \, ,
\end{eqnarray}
where the central value corresponds to $M_B^2 = 1.8$~GeV$^2$ and $s_0 = 6.0$~GeV$^2$, and the uncertainties are due to the Borel mass $M_B$, the threshold value $s_0$, and various condensates listed in Eqs.~(\ref{condensates}), respectively.

%
%=====================================================================================
%=====================================================================================
\section{Mixing of Currents}\label{sec:mixing}
%=====================================================================================
%=====================================================================================
%

In the previous section we have used the two single $s s \bar s \bar s$ tetraquark currents with $J^{PC} = 1^{+-}$,  $\eta_{1\mu}$ and $\eta_{2\mu}$, to perform QCD sum rule analyses. In this section we further study their mixing. We shall follow the procedures used in Ref.~\cite{Chen:2018kuu}, where the mixing of two $s s \bar s \bar s$ tetraquark currents with $J^{PC} = 1^{--}$ is carefully investigated.

\begin{figure*}[hbt]
\begin{center}
\includegraphics[width=0.4\textwidth]{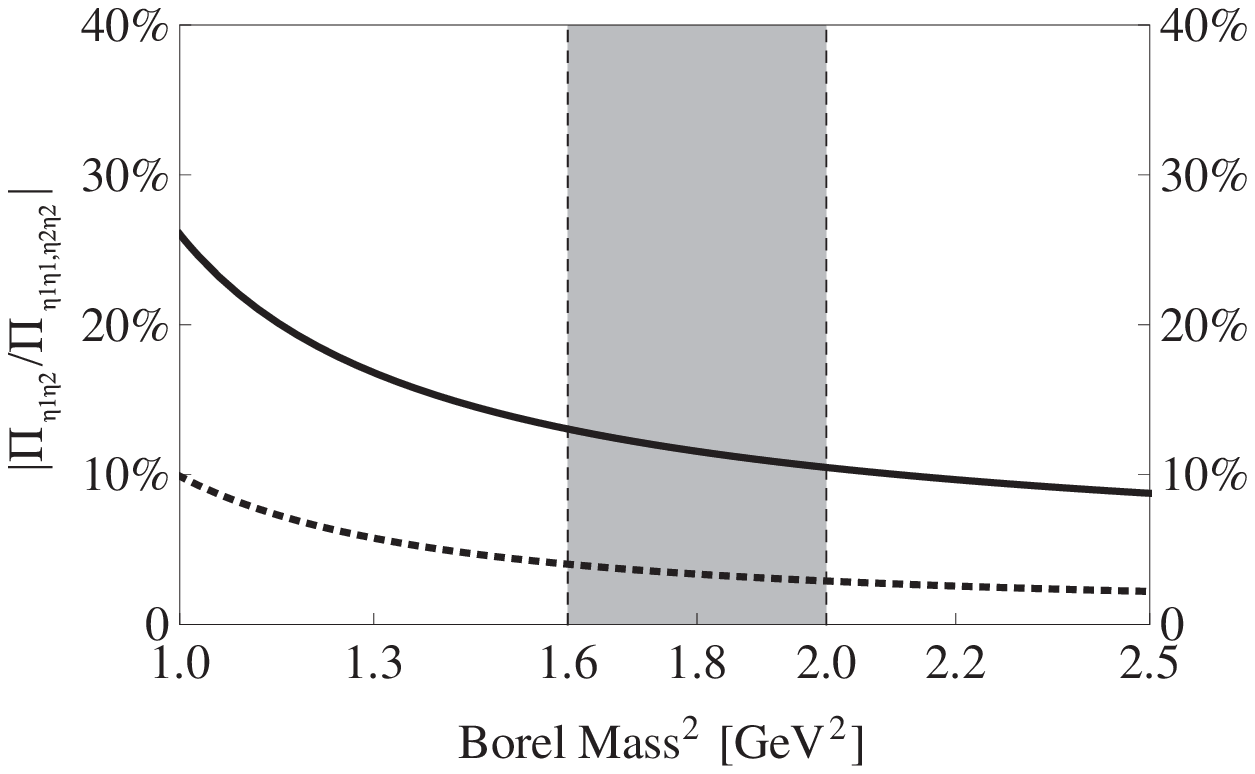}
\includegraphics[width=0.4\textwidth]{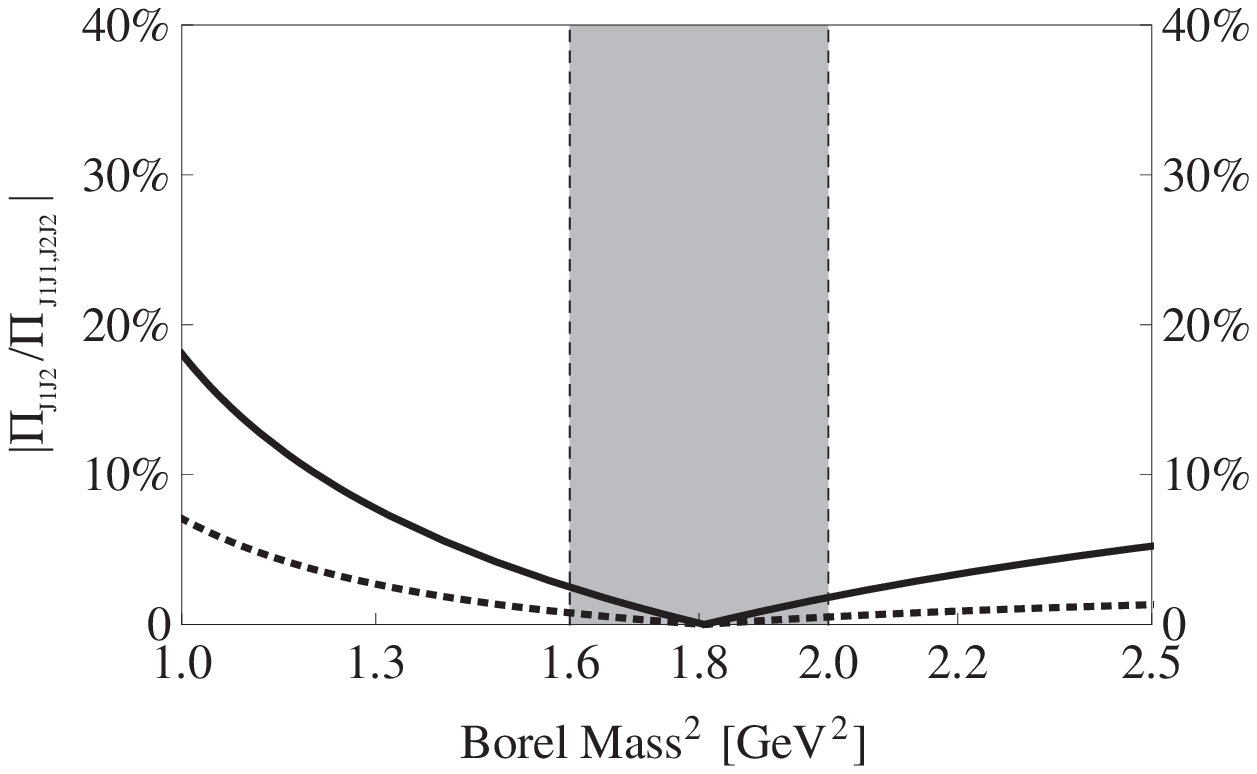}
\end{center}
\caption{In the left panel we show $\left|{\Pi_{\eta_1\eta_2}(M_B^2)\over\Pi_{\eta_1\eta_1}(M_B^2)}\right|$ (solid) and $\left|{\Pi_{\eta_1\eta_2}(M_B^2)\over\Pi_{\eta_2\eta_2}(M_B^2)}\right|$ (dotted) as functions of the Borel mass $M_B$, and in the right panel we show $\left|{\Pi_{J_1J_2}(M_B^2)\over\Pi_{J_1J_1}(M_B^2)}\right|$ (solid) and $\left|{\Pi_{J_1J_2}(M_B^2)\over\Pi_{J_2J_2}(M_B^2)}\right|$ (dotted) as functions of the Borel mass $M_B$. All curves are obtained by taking $s_0 = 6.0$ GeV$^2$.
}
\label{fig:offdiagonal}
\end{figure*}

To do this, first let us examine how large is the overlap between $\eta_{1\mu}$ and $\eta_{2\mu}$. We show the off-diagonal term $\Pi_{\eta_1\eta_2}(M_B^2)$ in the left panel of Fig.~\ref{fig:offdiagonal} as a function of the Borel mass $M_B$, compared with $\Pi_{\eta_1\eta_1}(M_B^2)$ and $\Pi_{\eta_2\eta_2}(M_B^2)$. This term has been defined in Eq.~(\ref{eq:offdiagonal}) and its explicit expression has been given in Eq.~(\ref{eq:eta12}). From these figures, it is difficult to judge whether the off-diagonal term is important or not, because it is neither too large nor too small. Hence, we further diagonalize the following matrix at around $M_B^2= 1.8$ GeV$^2$ and $s_0= 6.0$ GeV$^2$
\begin{eqnarray}
\left( \begin{array}{cc}
\Pi_{\eta_1\eta_1} & \Pi_{\eta_1\eta_2}
\\ \Pi_{\eta_1\eta_2}^\dagger & \Pi_{\eta_2\eta_2}
\end{array} \right) \, .
\end{eqnarray}
Then we obtain the mixing angle $\theta = 2.7^{\rm o}$ and two new currents $J_{1\mu}$ and $J_{2\mu}$ defined as:
\begin{eqnarray}
J_{1\mu} &=& \cos\theta~\eta_{1\mu} + \sin\theta~i~\eta_{2\mu} \, ,
\\ \nonumber J_{2\mu} &=& \sin\theta~\eta_{1\mu} + \cos\theta~i~\eta_{2\mu} \, .
\end{eqnarray}
These two new currents do not strongly correlate to each other in the region $1.6$~GeV$^2 < M_B^2< 2.0$~GeV$^2$, as shown in the right panel of Fig.~\ref{fig:offdiagonal}.

We use $J_{1\mu}$ and $J_{2\mu}$ to perform QCD sum rule analyses, and the results obtained are almost the same as those extracted from $\eta_{1\mu}$ and $\eta_{2\mu}$: a) $J_{1\mu}$ does not lead to reliable QCD sum rule results because $\Pi_{J_1J_1}(M_B^2, s_0)$ is negative in the region $s_0< 10$~GeV$^2$, and b) the mass extracted from $J_{2\mu}$ is about 2.00 GeV, the same as the one extracted from $\eta_{2\mu}$.

%
%=====================================================================================
%=====================================================================================
\section{Summary and Discussions}\label{sec:summary}
%=====================================================================================
%=====================================================================================
%

In this work we systematically construct all the $s s \bar s \bar s$ tetraquark currents with the quantum numbers $J^{PC} = 1^{+-}$. We find there are two independent ones ($\eta_{1\mu}$ and $\eta_{2\mu}$), which are then used to perform QCD sum rule analyses. The sum rules extracted from $\eta_{1\mu}$ and $\eta_{2\mu}$ are much different from each other: a) $\eta_{1\mu}$ does not lead to reliable results because $\Pi_{\eta_1\eta_1}(M_B^2, s_0)$ is negative, and so non-physical, in the region $s_0< 10$~GeV$^2$, and b) $\eta_{2\mu}$ leads to reliable results and the mass is extracted to be $2.00^{+0.10}_{-0.09}$~GeV, consistent with the second mass value listed in Eq.~(\ref{eq:mass2}), $2062.8 \pm 13.1 \pm 4.2$~{\rm MeV}. The mixing between $\eta_{1\mu}$ and $\eta_{2\mu}$ has been taken into account, and the results are the same. Hence, our results suggest that the structure $X$ observed at BESIII~\cite{BESIII} has the spin-parity quantum numbers $J^P = 1^{+-}$, and it can be interpreted as an $s s \bar s \bar s$ tetraquark state.

Recalling that in Refs.~\cite{Chen:2008ej,Chen:2018kuu} we have systematically investigated the $s s \bar s \bar s$ tetraquark states with $J^{PC} = 1^{--}$. There we also found two independent $s s \bar s \bar s$ tetraquark currents with $J^{PC} = 1^{--}$, but they lead to similar sum rule results, {\it i.e.,} the masses are extracted to be $2.34 \pm 0.17$ GeV and $2.41 \pm 0.25$ GeV, not far from each other~\cite{Chen:2018kuu}. These two values are both larger than the first mass value listed in Eq.~(\ref{eq:mass1}), $2002.1 \pm 27.5 \pm 15.0$~{\rm MeV}, suggesting that the structure $X$ observed at BESIII~\cite{BESIII} is difficult to be interpreted as an $s s \bar s \bar s$ tetraquark state of $J^{PC} = 1^{--}$.

\begin{table*}[hbt]
\begin{center}
\renewcommand{\arraystretch}{1.5}
\caption{Masses extracted from the vector and axial-vector tetraquark currents. Possible experimental candidates are listed for comparisons. We use $q$ to denote an up or down quark, and $s$ to denote a strange quark. The mass value $2.00^{+0.10}_{-0.09}$ GeV denoted by $^\dagger$ is obtained in the present study.}
\begin{tabular}{c | c c | c c | c c | c c }
\hline\hline
Contents & \multicolumn{2}{c|}{$J^{PC} = 1^{+-}$} & \multicolumn{2}{c|}{$J^{PC} = 1^{--}$} & \multicolumn{2}{c|}{$J^{PC} = 1^{++}$} & \multicolumn{2}{c}{$J^{PC} = 1^{-+}$}
\\ \cline{2-9}
(Isospin) & Theo. (GeV) & Exp. & Theo. (GeV) & Exp. & Theo. (GeV) & Exp. & Theo. (GeV) & Exp.
\\ \hline\hline
$q q \bar q \bar q$         & \multirow{4}{*}{1.47-1.66~\cite{Chen:2013jra}} & \multirow{4}{*}{--} & \multirow{2}{*}{1.60-1.73~\cite{Chen:2013jra}} & $\rho(1570)$~\cite{pdg} & \multirow{4}{*}{1.51-1.63~\cite{Chen:2013jra}} &
\multirow{4}{*}{$\begin{array}{c}a_1(1640)~\mbox{\cite{pdg}} \\ a_1(1420)~\mbox{\cite{Adolph:2015pws}} \end{array}$} & \multirow{2}{*}{$\sim1.6$~\cite{Chen:2008qw}} & \multirow{2}{*}{$\pi_1(1600)$~\cite{Adams:1998ff}}
\\ $(I = 1)$        &                            &                                             & &      $\rho(1700)$~\cite{pdg} &                              &                    &                     &
\\ \cline{1-1} \cline{4-5} \cline{8-9}
$q s \bar q \bar s$ &                            &                                                & \multirow{2}{*}{1.91-2.13~\cite{Chen:2013jra}} & $\rho(1900)$~\cite{pdg} & &
& \multirow{2}{*}{$\sim2.0$~\cite{Chen:2008qw}} & \multirow{2}{*}{$\pi_1(2015)$~\cite{Kuhn:2004en}}
\\ $(I = 1)$        &                            &                                               & & $\rho(2150)$~\cite{pdg} &                              &                    &                     &
\\ \hline
$s s \bar s \bar s$ & \multirow{2}{*}{$2.00 ^{+0.10}_{-0.09}~^\dagger$} & \multirow{2}{*}{$X(2063)$~\cite{BESIII}} & $2.34 \pm 0.17$~\cite{Chen:2018kuu} & $Y(2175)$~\cite{Aubert:2006bu}  & \multirow{2}{*}{--} & \multirow{2}{*}{--} & \multirow{2}{*}{--} & \multirow{2}{*}{--}
\\ \cline{4-5}
(I = 0)             &                            &                                  & $2.41 \pm 0.25$~\cite{Chen:2018kuu}& $Y(2470)$~\cite{Aubert:2007ur}  &                    &                     &                    &
\\ \hline \hline
\end{tabular}
\label{tab:results}
\end{center}
\end{table*}

Besides these isoscalar $s s \bar s \bar s$ tetraquark states, in Ref.~\cite{Chen:2013jra} we have systematically constructed all the isovector tetraquark currents of $I^GJ^{PC} = 1^+1^{+-}/1^+1^{--}/1^-1^{++}/1^-1^{-+}$, and found a one-to-one correspondence among them, {\it i.e.,} for every tetraquark current of $I^GJ^{PC} = 1^+1^{+-}$ one can construct a corresponding one of $I^GJ^{PC} = 1^+1^{--}$, etc. These tetraquark currents have been used to perform QCD sum rule analyses in Refs.~\cite{Chen:2008qw,Chen:2013jra}, and the results are summarized in Table~\ref{tab:results}, where $q$ denotes an up or down quark, and $s$ denotes a strange quark. Note that the sum rule results do not have the above one-to-one correspondence, for examples: a) there are four $q q \bar q \bar q$ currents and four $q s \bar q \bar s$ currents with $I^GJ^{PC} = 1^+1^{+-}$, and the masses extracted from these currents are all around 1.47-1.66 GeV; b) there are also four $q q \bar q \bar q$ currents and four $q s \bar q \bar s$ currents with $I^GJ^{PC} = 1^+1^{--}$, but the masses extracted from the former four are around 1.60-1.73 GeV and the masses extracted from the latter four are around 1.91-2.13 GeV. This behaviour may relate to their internal structures, such as internal orbital excitations.

Similarly, there is a one-to-one correspondence among the $s s \bar s \bar s$ tetraquark currents with $J^{PC} = 1^{+-}/1^{--}/1^{++}/1^{-+}$. Those with $J^{PC} = 1^{+-}$ and $1^{--}$ have been used to perform QCD sum rule analyses in the present study as well as in Refs.~\cite{Chen:2008ej,Chen:2018kuu}. The results are also summarized in Table~\ref{tab:results}. From this table, we propose to search for the $s s \bar s \bar s$ tetraquark states with $J^{PC} = 1^{++}$ and $1^{-+}$ in future experiments. We are now studying them following the same approach used in the present study. Their masses may also be around 2.0-2.4 GeV, and the possible decay channels to observe them are $\eta^\prime f_0(980)$, $\eta^\prime K \bar K$, and $\eta^\prime K \bar K^*$, etc.

When studying light tetraquark states, it is usually difficult to determine the experimental signal as a genuine four-quark state other than a conventional $\bar q q$ meson, because the signal always has a quite large decay width. For example, besides the $s s \bar s \bar s$ tetraquark state of $J^{PC} = 1^{+-}$, there are many other possible interpretations to explain the structure $X$, such as the second radial excitation of $h_1(1380)$ having $I(J^P) = 0(1^+)$~\cite{liu}. However, with the large amount of data collected at BESIII, this problem may be partly solved, and it is promising to continuously study light exotic hadrons. Together with those studies on charmonium-like $XYZ$ states, our understudying on the nature of exotic hadrons can be significantly improved.

%
%=====================================================================================
%=====================================================================================
%=====================================================================================
\section*{Acknowledgments}
%=====================================================================================
%=====================================================================================
%=====================================================================================
%

We thank Professor Shi-Lin Zhu for useful discussions.
This project is supported by
the National Natural Science Foundation of China under Grants No. 11575017, No. 11722540, and No. 11761141009,
the Fundamental Research Funds for the Central Universities,
and the Chinese National Youth Thousand Talents Program.

%
%%%%%%%%%%%%%%%%%%%%%%%%%%%%%%%%%%%%%%%%%%%%%%%%%%%%%%%%%%%%%%%%%%%%%%%%%%%%%%

%%%%%%%%%%%%%%%%%%%%%%%%%%%%%%%%%%%%%%%%%%%%%%%%%%%%%%%%%%%%%%%%%%%%%%%%%%%%%%
%

\end{document}